\documentclass[10pt, prb, aps, twocolumn, showpacs, citeautoscript, floatfix, reprint, amsmath, amssymb, notitlepage, superscriptaddress, longbibliography]{revtex4-1}

\usepackage{graphicx}
\usepackage{stmaryrd}
\usepackage{rotating}
\usepackage{amsmath}
\usepackage{amsfonts}
\usepackage{amssymb}
\usepackage{wasysym}
\usepackage[countmax]{subfloat}
\usepackage{dcolumn} 
\usepackage{bm} 
\usepackage{color}

\usepackage{float}
\usepackage{latexsym}

\begin{document}

\title{Absorbance marker: Detection of quantum geometry and spread of Wannier function in disordered 2D semiconductors}

\author{Luis F. C\'{a}rdenas-Castillo}

\affiliation{Department of Physics, PUC-Rio, 22451-900 Rio de Janeiro, Brazil}

\author{Shuai Zhang}
\email{zhangshuai0018@outlook.com}
\affiliation{Department of Physics, PUC-Rio, 22451-900 Rio de Janeiro, Brazil}

\author{Fernando L. Freire Jr.}

\affiliation{Department of Physics, PUC-Rio, 22451-900 Rio de Janeiro, Brazil}

\author{Wei Chen}
\email{wchen@puc-rio.br}
\affiliation{Department of Physics, PUC-Rio, 22451-900 Rio de Janeiro, Brazil}

\date{\rm\today}

\begin{abstract}

The optical absorbance of 2D semiconductors is generalized to individual lattice sites through the topological marker formalism, yielding an absorbance marker. This marker allows to investigate the atomic scale variation of absorbance caused by impurities, thereby quantifies the influence of disorder on the quantum geometry and the spread of Wannier functions of valence band states. Applying this marker to transition metal dichalcogenides reveals a very localized suppression of absorbance caused by potential impurities, rendering a reduction of absorbance in the macroscopic scale proportional to the impurity density, in good agreement with the experimental results of plasma-treated WS$_{2}$.

\end{abstract}

\maketitle

\section{Introduction}

Ever since the discovery of graphene, the research in 2D materials has blossomed into a major area in condensed matter physics with promising applications\cite{Novoselov04,Geim07,CastroNeto09}. The flexibility of 2D materials, such as the possibility to stack and twist them, makes them an ideal playground to incorporate strong correlations and band structure effects. Many new phenomena have been discovered owing to this flexibility, such as the flatbands and superconductivity in twisted bilayer graphene\cite{Cao18,Cao18_2}, which have igniting a great deal of interest in 2D materials recently. Turning to the optical responses of 2D semiconductors, as we shall focus in the present work, the optical properties of these materials are equally exciting, with many intriguing features that may even be perceived by naked eyes in the macroscopic scale. For instance, the roughly frequency-independent visual opacity of monolayer graphene is known to be given by $\pi$ times the fine-structure constant\cite{Nair08,Nair10,Stauber08} $\pi\alpha\approx 2.3\%$, which has been argued to be a manifestation of topological charge that can be perceived in the macroscopic scale\cite{deSousa23_graphene_opacity}.

Recent theoretical advances on 2D semiconductors have further fueled the motivation to investigate their optical properties, especially the clarification of the connection between optical absorbance and the surging notion of quantum geometry. Theoretically, the fidelity of two valence band Bloch states at slightly different momenta defines a quantum metric\cite{Provost80}, through which many differential geometrical properties like Riemann tensor and Euler characteristic can be introduced, giving rise to the notion of quantum geometry\cite{Matsuura10,Ma13,Ma14,Kolodrubetz13,Kolodrubetz17,Smith22,Chen25_quantum_geometry_topological}. Because this quantum metric is equivalently the matrix element of interband optical transition, the absorbance of 2D semiconductors can be regarded as a quantum geometric effect, leading to many remarkable consequences. For instance, the aforementioned connection between the $\pi\alpha$ absorbance of graphene and the topological charge is because of a metric-curvature correspondence between the quantum metric and topological order\cite{Ma13,Ma14,Yang15,Piechon16,Panahiyan20_fidelity_susceptibility,vonGersdorff21_metric_curvature,Mera22}. Furthermore, it is recently proposed that the absorbance divided by frequency and then integrated over frequency gives the momentum integration of quantum metric\cite{CardenasCastillo24_Wannier_opacity}. The significance of this optical sum rule is that it offers a direct experimental measurement for the spread of valence band Wannier function, a quantity that has been heavily exploited in density functional theory. Using hexagonal transition metal dichalcogenides MX$_2$ (M = Mo, W; X = S, Se, Te) as examples, the theoretical calculation using a three-band tight-binding model\cite{Liu13} yields an excellent agreement with the absorbance measured experimentally\cite{Li14,Li17}, and the spread of the $d$-orbital Wannier function of the transition metal has been extracted\cite{CardenasCastillo24_Wannier_opacity}, demonstrating the feasibility of this optical sum rule.

In this paper, we advance the notion of quantum geometry in 2D semiconductors by presenting a formalism that maps the absorbance to individual lattice sites, yielding what we call the absorbance marker. This mapping is achieved through adopting the formalism of topological markers that maps the topological order to lattice sites, which was originally proposed for 2D time-reversal breaking Chern insulators\cite{Prodan10,Prodan11,Bianco11}, but recently generalized to any dimension and symmetry class\cite{Chen23_universal_marker}. The advantage of this absorbance marker is that it clarifies the experimental protocol of measuring the momentum integration of quantum metric and the spread of Wannier function\cite{Marzari97,Marzari12} generalized to disordered systems, which have been called localization marker or fidelity marker\cite{Marrazzo19,deSousa23_fidelity_marker}, allowing the effect of disorder on quantum geometry and Wannier orbitals to be quantified. Using TMD materials as concrete examples, we reveal that the absorbance is reduced at the impurity sites that have a positive impurity potential, intuitively attributed to the reduced number of electrons available to absorb light. The reduction is very local in space, i.e., almost entirely confined to the impurity sites, and consequently causing a very local reduction of the spread of Wannier orbitals. On the other hand, the absorbance measured in the macroscopic scale is the spatial average of that on individual lattice sites, so the experimental absorbance should only slightly reduce in disordered samples, with more reduction at higher impurity density. This theoretical prediction is nicely reproduced by our experimental results of plasma-treated WS$_{2}$, which show more and more reduction of absorbance as the sample is subject to longer and longer plasma exposure, which presumably removes the electrons on tungsten sites. Thus our absorbance marker not only serves as a theoretical tool to explain the absorbance of disordered 2D semiconductors in the macroscopic scale, but also describes how the optical responses and quantum geometric properties change in the atomic scale, and is able to incorporate any type of semiconducting band structure and microscopic detail of the impurities.


\section{Absorbance marker for 2D materials}

\subsection{Absorbance marker \label{sec:absorbance_marker_formalism}}

We begin by reviewing the concept of quantum geometry applying to homogeneous 2D semiconductors, and elaborating how it is related to the absorbance and the spread of Wannier function. Consider a fully gapped 2D semiconductor or insulator that contains $N_{-}$ valence bands, whose fully antisymmetric many-body valence band Bloch state at momentum ${\bf k}=(k_{x},k_{y})$ is described by
\begin{eqnarray}
|u^{\rm val}({\bf k})\rangle=\frac{1}{\sqrt{N_{-}!}}\varepsilon^{n_{1}n_{2}...n_{N-}}|n_{1}\rangle|n_{2}\rangle...|n_{N_{-}}\rangle.\;\;\;
\label{psi_val}
\end{eqnarray}
where $\varepsilon^{n_{1}n_{2}...n_{N-}}$ is the Levi-Civita symbol, and $|n_{i}\rangle$ denotes each valence band eigenstate. As we will demonstrate below, the absorbance is closely related to the quantum metric of this fully antisymmetric valence band state, defined from the overlap of two such states at neighboring momenta\cite{Provost80} 
\begin{eqnarray}
|\langle u^{\rm val}({\bf k})|u^{\rm val}({\bf k+\delta k})\rangle|=1-\frac{1}{2}g_{\mu\nu}({\bf k})\delta k^{\mu}\delta k^{\nu},
\label{uval_gmunu}
\end{eqnarray}
that is known to be related to each valence band state $|n\rangle$ and conduction band state $|m\rangle$ by\cite{Matsuura10,vonGersdorff21_metric_curvature} 
\begin{eqnarray}
&&g_{\mu\nu}({\bf k})=\frac{1}{2}\langle \partial_{\mu}u^{\rm val}|\partial_{\nu}u^{\rm val}\rangle+\frac{1}{2}\langle \partial_{\nu}u^{\rm val}|\partial_{\mu}u^{\rm val}\rangle
\nonumber \\
&&-\langle \partial_{\mu}u^{\rm val}|u^{\rm val}\rangle \langle u^{\rm val}|\partial_{\nu}u^{\rm val}\rangle
\nonumber \\
&&=
\frac{1}{2}\sum_{nm}\left[\langle \partial_{\mu}n|m\rangle\langle m|\partial_{\nu}n\rangle+\langle \partial_{\nu}n|m\rangle\langle m|\partial_{\mu}n\rangle\right],
\label{gmunu_T0}
\end{eqnarray}
with $\partial_{\mu}\equiv\partial/\partial k^{\mu}$. To relate this quantum metric to optical responses at a specific frequency, we introduce a dressed quantum metric spectral function that turns out to be proportional to the optical conductivity\cite{Chen22_dressed_Berry_metric} 
\begin{eqnarray}
&&g_{\mu\mu}^{d}({\bf k},\omega)=\sum_{\ell<\ell '}\langle\partial_{\mu}\ell|\ell '\rangle\langle\ell '|\partial_{\mu}\ell\rangle
\nonumber \\
&&\times\left[f(\varepsilon_{\ell}^{\bf k})-f(\varepsilon_{\ell '}^{\bf k})\right]\delta(\omega+\frac{\varepsilon_{\ell}^{\bf k}}{\hbar}-\frac{\varepsilon_{\ell '}^{\bf k}}{\hbar})
\nonumber \\
&&=\frac{A_{\rm cell}}{\pi e^{2}\hbar\omega}\sigma_{\mu\mu}({\bf k},\omega),
\label{gmunuw_finiteT_general}
\end{eqnarray}
where $A_{\rm cell}$ is the area of the unit cell, and $\sigma_{\mu\mu}({\bf k},\omega)$ is the finite temperature longitudinal optical conductivity at momentum ${\bf k}$ and frequency $\omega$. The optical conductivity measured in the macroscopic scale is given by the momentum integration of that in the momentum space
\begin{eqnarray}
&&\sigma_{\mu\mu}(\omega)=A_{\rm cell}\int\frac{d^{2}{\bf k}}{(2\pi\hbar)^{2}}\,\sigma_{\mu\mu}({\bf k},\omega)
\nonumber \\
&&=\frac{\pi e^{2}}{\hbar}\,\omega\int\frac{d^{2}{\bf k}}{(2\pi)^{2}}\,g_{\mu\mu}^{d}({\bf k},\omega).
\label{fidelity_number_spec_fn}
\end{eqnarray}
On the other hand, the incident power per unit cell of the light is $W_{i}=c\varepsilon_{0}E_{0}^{2}/2$, and the absorption power of the material is $W_{a}=\sigma_{\mu\mu}(\omega)E_{0}^{2}/2$, where $E_{0}$ is the strength of the electric field. The absorbance at polarization $\mu$ and frequency $\omega$ is then given by\cite{Nair08,Stauber08}
\begin{eqnarray}
&&{\cal O}(\omega)=\frac{W_{a}^{\mu}(\omega)}{W_{i}}.
\label{Ow_graphene_finiteT}
\end{eqnarray}
If the incident light is unpolarized, then the real part of optical conductivity relevant to the absorbance is the average of that in the two crystalline directions, usually called $\sigma_{1}(\omega)$
\begin{eqnarray}
\sigma_{1}(\omega)=\frac{1}{2}\left[\sigma_{xx}(\omega)+\sigma_{yy}(\omega)\right].
\label{sigma1_average_sigmamu}
\end{eqnarray}
Thus we may write the absorbance of any 2D semiconductor as
\begin{eqnarray}
{\cal O}(\omega)=\frac{\sigma_{1}(\omega)}{c\varepsilon_{0}}
=4\times\pi\alpha\times\frac{\sigma_{1}(\omega)}{e^{2}/\hbar},
\label{absorbance_vs_sigma1}
\end{eqnarray}
which describes the frequency dependence of the absorbance of homogeneous 2D semiconductors where the momentum ${\bf k}$ remains a good quantum number.

We remark that strictly speaking, the interband transition described by Eq.~(\ref{gmunuw_finiteT_general}) is not valid for the optical conductivity in indirect band gap semiconductors at low frequency. For indirect band gap semiconductors, the optical absorption at low frequency involves the valence band state $|n({\bf k}_{1})\rangle$ and conduction band state $|n({\bf k}_{2})\rangle$ at different momenta ${\bf k}_{1}\neq{\bf k}_{2}$, as well as the emission of phonons, so the absorption is not simply determined by the quantum metric at a specific momentum. However, this low frequency absorption is usually very small in practice\cite{Yu10}, and does not contribute to the spread of Wannier function and the momentum integration of quantum metric that will be introduced in the following sections. We will assume that the indirect band gap absorption is small and ignore them throughout the discussion.

Our goal in the present work is to generalize this frequency-dependent absorbance to a local quantity that can be defined on individual lattice sites, such that the effect of disorder can be investigated. For this purpose, we resort to the formalism of topological markers that maps the topological order to individual lattice sites\cite{Prodan10,Prodan11,Bianco11}. In fact, such a formalism has been applied to describe the atomic scale dielectric and optical properties of 3D semiconductors, yielding the dielectric and optical markers\cite{Chen25_optical_marker}. The optical conductivity marker therein can be straightforwardly generalized to 2D semiconductors by simply replacing the positions on a 3D lattice sites ${\bf r}=(x,y,z)$ by that defined on 2D lattice sites ${\bf r}=(x,y)$, yielding
\begin{widetext}
\begin{eqnarray}
&&\sigma_{\mu\mu}({\bf r},\omega)=\left(\frac{\pi e^{2}}{\hbar A_{\rm cell}}\right)\omega\sum_{s}\langle{\bf r},s|\left\{\sum_{\ell<\ell '}S_{\ell}{\hat\mu}S_{\ell '}{\hat\mu}S_{\ell}[f(E_{\ell})-f(E_{\ell '})]\delta\left(\omega+\frac{E_{\ell}}{\hbar}
-\frac{E_{\ell '}}{\hbar}\right)\right\}|{\bf r},s\rangle.
\end{eqnarray}
where $S_{\ell}=|E_{\ell}\rangle\langle E_{\ell}|$ is the projector to an eigenstate $|E_{\ell}\rangle$ of the lattice Hamiltonian $H$ satisfying $H|E_{\ell}\rangle=E_{\ell}|E_{\ell}\rangle$, ${\hat\mu}=\left\{{\hat x},\hat{y}\right\}$ is the position operator along the $\mu$-direction, and $s$ enumerates all the degrees of freedom inside a unit cell. Inserting this expression into Eqs.~(\ref{sigma1_average_sigmamu}) and (\ref{absorbance_vs_sigma1}), we can introduce an absorbance marker by
\begin{eqnarray}
{\cal O}({\bf r},\omega)
=\left(\frac{2\pi^{2}\alpha}{A_{\rm cell}}\right)\omega\sum_{\mu}\sum_{s}\langle{\bf r},s|\left\{\sum_{\ell<\ell '}S_{\ell}{\hat\mu}S_{\ell '}{\hat\mu}S_{\ell}[f(E_{\ell})-f(E_{\ell '})]\delta\left(\omega+\frac{E_{\ell}}{\hbar}
-\frac{E_{\ell '}}{\hbar}\right)\right\}|{\bf r},s\rangle.
\label{absorbance_marker_definition}
\end{eqnarray}
Numerically, we implement the optical conductivity marker and absorbance marker by introducing the operator
\begin{eqnarray}
&&{\cal M}_{\mu}(\omega)=\sum_{\ell<\ell '}S_{\ell}{\hat\mu}S_{\ell '}\sqrt{[f(E_{\ell})-f(E_{\ell '})]\delta\left(\omega+\frac{E_{\ell}}{\hbar}
-\frac{E_{\ell '}}{\hbar}\right)},
\label{Mw_operator}
\end{eqnarray}
\end{widetext}
and express the markers by
\begin{eqnarray}
&&\sigma_{\mu\mu}({\bf r},\omega)=\left(\frac{\pi e^{2}}{\hbar A_{\rm cell}}\right)\omega\sum_{s}\langle{\bf r},s|{\cal M}_{\mu}(\omega){\cal M}_{\mu}^{\dag}(\omega)|{\bf r},s\rangle,
\nonumber \\
&&{\cal O}({\bf r},\omega)
=\left(\frac{2\pi^{2}\alpha}{A_{\rm cell}}\right)\omega\sum_{\mu}\sum_{s}\langle{\bf r},s|{\cal M}_{\mu}(\omega){\cal M}_{\mu}^{\dag}(\omega)|{\bf r},s\rangle,\;\;\;\;\;\;
\end{eqnarray}
which serves as a convenient numerical recipe to implement the markers.

We now comment on the choice of position operator $\hat{\mu}$. For a rectangular lattice of with $L$ unit cells in each direction, the common choice of the position operator, say along the ${\bf x}$ direction, is 
\begin{eqnarray}
{\hat x}={\rm diag}\left(1,2,3...L\right)\otimes I_{s},
\label{position_operator_1234}
\end{eqnarray}
where $I_{s}$ is a $s\times s$ identity matrix, indicating that all degrees of freedom $s$ within the same unit cell are designated with the same position operator. However, this position operator exhibits a sharp jump at the boundary from $L$ to $1$, causing an anomaly on the boundary sites even if periodic boundary condition is imposed\cite{Bianco11}. To fixed this problem, a position operator that respects the translational invariance has been proposed
\begin{eqnarray}
{\hat x}=\frac{L}{2\pi}{\rm diag}\left(e^{2\pi i/L},e^{4\pi i/L},e^{6\pi i/L}...e^{2\pi i}\right)\otimes I_{\sigma},\;\;\;
\label{position_operator_exp}
\end{eqnarray}
and likewisely for ${\hat y}$, i.e., using the enumeration $\frac{L}{2\pi}\exp(2\pi ix/L)$ in the position operator\cite{Aligia99,Prodan10,Aligia23,Molignini23_TPT_finite_T,Oliveira24_robustness_marker}. This exponentiated position operator can be easily generalized to other 2D lattices, such as the triangular lattice for TMD materials examined in the following sections. The resulting absorbance marker is complex, and its modulus gives the correct value of homogeneous absorbance on every lattice site. For the numerically results presented in the following section, we use this exponentiated position operator and simply denote the modulus of the markers as the desired markers
\begin{eqnarray}
&&\sigma_{\mu\mu}({\bf r},\omega)\rightarrow\left[({\rm Re}\,\sigma_{\mu\mu}({\bf r},\omega))^2+({\rm Im}\,\sigma_{\mu\mu}({\bf r},\omega))^2\right]^{1/2},
\nonumber \\
&&{\cal O}({\bf r},\omega)\rightarrow\left[({\rm Re}\,{\cal O}({\bf r},\omega))^2+({\rm Im}\,{\cal O}({\bf r},\omega))^2\right]^{1/2}.
\label{modulus_Cr}
\end{eqnarray}
Using this method, we can reliably investigate the effect of disorder on the local absorbance, as we shall see in the following sections.

\subsection{Detection of quantum geometric properties in disordered 2D semiconductors \label{sec:detection_quantum_geometry}}

The absorbance marker introduced in the previous section allows to quantify the influence of disorder on the momentum space quantum geometry of the fully antisymmetric valence band state. As a result, it also quantifies the influence of disorder on the gauge-invariant part of the spread of valence-band Wannier functions, as we now elaborate. For our purpose, we first review several quantum geometric quantities in homogeneous 2D semiconductors. Firstly, the frequency integration of the quantum metric spectral function in Eq.~(\ref{gmunuw_finiteT_general}) gives the dressed quantum metric\cite{Chen22_dressed_Berry_metric}, which may be viewed as a finite temperature generalization of the quantum metric
\begin{eqnarray}
g_{\mu\mu}^{d}({\bf k})=\int_{0}^{\infty}d\omega\,g_{\mu\mu}^{d}({\bf k},\omega).
\end{eqnarray}
The momentum integration of this quantity further yields the diagonal elements of what we call the fidelity number\cite{deSousa23_fidelity_marker} 
\begin{eqnarray}
{\cal G}_{\mu\mu}^{d}=\int\frac{d^{2}{\bf k}}{(2\pi)^{2}}g_{\mu\mu}^{d}({\bf k}),
\label{fidelity_number_definition}
\end{eqnarray}
which at zero temperature characterizes the average distance between neighboring Bloch states in the BZ. Furthermore, for a valence-band Bloch state $|n\rangle$, the corresponding Wannier state is given by
\begin{eqnarray}
&&|{\bf R} n\rangle=\sum_{\bf k}e^{i {\bf k}\cdot({\hat{\bf r}}-{\bf R})/\hbar}|n\rangle,\;\;\;\;\;
\label{Wannier_basis}
\end{eqnarray}
with ${\bf r}$ and ${\bf R}$ the position and Bravais lattice vectors, respectively. The $\langle {\bf r}|{\bf R} n\rangle=W_{n}({\bf r}-{\bf R})$ gives the Wannier function of the $n$th-band located around the unit cell at ${\bf R}$. One may further introduce the gauge-invariant part of the spread of valence-band Wannier function by\cite{Marzari97,Marzari12,Souza08,CardenasCastillo24_Wannier_opacity}
\begin{eqnarray}
&&\Omega_{I}=\sum_{n}\left[\langle{\bf 0}n|r^{2}|{\bf 0}n\rangle-\sum_{{\bf R}n'}|\langle{\bf R}n'|{\bf r}|{\bf 0}n\rangle|^{2}\right]
\nonumber \\
&&=\lim_{T\rightarrow 0}A_{\rm cell}{\rm Tr}\,{\cal G}_{\mu\nu}^{d}
=\lim_{T\rightarrow 0}\frac{A_{\rm cell}}{2\pi}\int_{0}^{\infty}\frac{d\omega}{\omega}\left[\frac{{\cal O}(\omega)}{\pi\alpha}\right],
\label{OmegaI_Gmunu_Ow}
\end{eqnarray}
that is known to be given by the trace of the fidelity number in Eq.~(\ref{fidelity_number_definition}) at zero temperature, and is also given by the absorbance ${\cal O}(\omega)$ divided by frequency and then integrated over frequency according to the formalism in Sec.~\ref{sec:absorbance_marker_formalism}, where $A_{\rm cell}$ the area of the unit cell. This relation provides a concrete experimental protocol to measure the spread $\Omega_{I}$ and the trace ${\rm Tr}\,{\cal G}_{\mu\nu}^{d}$, as has been elaborated for homogeneous TMD materials\cite{CardenasCastillo24_Wannier_opacity}.

Our aim in the present work is to map the quantities $\left\{\Omega_{I},{\rm Tr}\,{\cal G}_{\mu\nu}^{d},{\cal O}(\omega)\right\}$ in Eq.~(\ref{OmegaI_Gmunu_Ow}) to every lattice site such that they can be defined locally even for disordered systems. The mapping of fidelity number to real space has been proposed previously, yielding what is called the localization marker\cite{Marrazzo19} or fidelity marker\cite{deSousa23_fidelity_marker} ${\rm Tr}\,{\cal G}_{\mu\nu}^{d}({\bf r})$. Introducing an operator that is analogous to that in Eq.~(\ref{Mw_operator})
\begin{eqnarray}
&&{\cal M}_{\mu}=\sum_{\ell<\ell '}S_{\ell}{\hat\mu}S_{\ell '}\sqrt{[f(E_{\ell})-f(E_{\ell '})]},
\end{eqnarray}
the matrix element of the fidelity marker takes the form 
\begin{eqnarray}
&&{\cal G}_{\mu\nu}^{d}({\bf r})=\frac{1}{2A_{\rm cell}}\sum_{s}{\rm Re}\left\{\langle{\bf r},s|\left[{\cal M}_{\mu}{\cal M}_{\nu}^{\dag}+{\cal M}_{\nu}{\cal M}_{\mu}^{\dag}\right]|{\bf r},s\rangle\right\}.
\nonumber \\
\label{finiteT_fidelity_marker_XY}
\end{eqnarray}
The mapping of absorbance to lattice sites gives precisely the absorbance marker introduced in Sec.~\ref{sec:absorbance_marker_formalism}, so these local quantities are related in a way that is completely analogous to Eq.~(\ref{OmegaI_Gmunu_Ow}) 
\begin{eqnarray}
&&\Omega_{I}({\bf r})\equiv\lim_{T\rightarrow 0}A_{\rm cell}{\rm Tr}\,{\cal G}_{\mu\nu}^{d}({\bf r})
\nonumber \\
&&=\lim_{T\rightarrow 0}\frac{A_{\rm cell}}{2\pi}\int_{0}^{\infty}\frac{d\omega}{\omega}\left[\frac{{\cal O}({\bf r},\omega)}{\pi\alpha}\right],
\label{OmegaI_Gmunu_Ow_atr}
\end{eqnarray}
which is readily applicable to any disordered 2D semiconductor. Note that both Bloch state and Wannier state are strictly speaking not well-defined in disordered systems, so one cannot express the quantity $\Omega_{I}({\bf r})$ in terms of Wannier states. Nevertheless, this quantity deined directly from the fidelity marker $\Omega_{I}({\bf r})\equiv\lim_{T\rightarrow 0}A_{\rm cell}{\rm Tr}\,{\cal G}_{\mu\nu}^{d}({\bf r})$ has been argued to be able to quantify the insulating nature of the material, thereby serving as a legitimate generalization of the spread $\Omega_{I}$ to disordered systems\cite{Marrazzo19}. Equation (\ref{OmegaI_Gmunu_Ow_atr}) then further points to the experimental detection of this local spread $\Omega_{I}({\bf r})$ by means of the local absorbance ${\cal O}({\bf r},\omega)$, which we anticipate may be measured by scanning thermal microscopy that can probe the atomic scale heat absorption under a specific frequency of light\cite{Majumdar99,Kittel05,Gomes15,Zhang20}.

Our absorbance marker formalism also points to a much simpler experiment to quantify the disorder effect: Simply measure the absorbance in the macroscopic scale, which corresponds to the spatial average of these local markers 
\begin{eqnarray}
&&\left\{\overline{\Omega_{I}},\overline{{\rm Tr}\,{\cal G}_{\mu\nu}^{d}},\overline{{\cal O}(\omega)}\right\}
\nonumber \\
&&=\frac{1}{N}\sum_{\bf r}\left\{\Omega_{I}({\bf r}),{\rm Tr}\,{\cal G}_{\mu\nu}^{d}({\bf r}),{\cal O}({\bf r},\omega)\right\},
\label{avg_OmegaI_TrG_O}
\end{eqnarray}
where $N$ is the number of lattice sites. Thus the experimental protocol given in Eq.~(\ref{OmegaI_Gmunu_Ow}), i.e., dividing the macroscopic absorbance $\overline{{\cal O}(\omega)}$ by frequency and then integrating over frequency, can be used to detect the spatially averaged spread $\overline{\Omega_{I}}$. Consequently, the trace of spatially averaged fidelity marker $\overline{{\rm Tr}\,{\cal G}_{\mu\nu}^{d}}=\overline{\Omega_{I}}/A_{\rm cell}$ can be readily measured, offering a concrete experimental protocol to quantify the quantum geometry under the influence of disorder\cite{Marrazzo19,deSousa23_fidelity_marker}. We will used disordered TMD materials in the following section to demonstrate the feasibility of our proposal. 


\section{Applications to transition metal dichalcogenides}

\subsection{Minimal tight-binding model for transition metal dichalcogenides \label{sec:tight-binding_model_TMD}}

We proceed to use hexagonal TMDs to demonstrate the absorbance marker. The model has chemical formula 1H-MX$_{2}$, with transition metal M = Mo, W, and chalcogen X = S, Se, Te. To simplify the calculation, we adopt the spinless version of a generic tight-binding Hamiltonian proposed by Ref.~\onlinecite{Liu13} without the inclusion of spin-orbit coupling, since the spin-orbit coupling is a very small energy scale and only has a very minor correction to the frequency-dependence of the absorbance ${\cal O}(\omega)$, yet the tight-binding model is much simpler to calculate. The three-band model is described by the basis $|\psi\rangle=(d_{z^{2}},d_{xy},d_{x^{2}-y^{2}})^{T}$ of the $d$ orbitals of the transition metal atoms (Mo,W), with a $3\times 3$ Hamiltonian in momentum space parametrized by
\begin{eqnarray}
H_{0}({\bf k})=\left(\begin{array}{ccc}
V_{0} & V_{1} & V_{2} \\
V_{1}^{\ast} & V_{11} & V_{12} \\
V_{2}^{\ast} & V_{12}^{\ast} & V_{22}
\end{array}\right)
\end{eqnarray}
where the ${\bf k}$-dependent functions $V_{0}\sim V_{22}$ and the parameters for each TMD compound can be found in Ref.~\onlinecite{Liu13}. The real space tight-binding Hamiltonian  can then be obtained by a straightforward Fourier transform of $H_{0}({\bf k})$, whose lengthy expression will be omitted. For the sake of comparison with experiments in the next section, we will use the parameters for WS$_{2}$, and consider an impurity potential that is assumed to be the same for all three orbitals
\begin{eqnarray}
H_{imp}=\sum_{i\in imp}\sum_{s}U_{imp}c_{is}^{\dag}c_{is},
\end{eqnarray}
where $s$ denotes the three orbitals on the impurity sites $i\in imp$, and $U_{imp}$ is the impurity potential. After the eigenstates and eigenenergies are found via diagonalization $H|E_{n}\rangle=E_{n}|E_{n}\rangle$ of the full Hamiltonian $H=H_{0}+H_{imp}$, we use them to calculate the absorbance marker ${\cal O}({\bf r},\omega)$ and the local spread $\Omega_{I}({\bf r})$ according to the formulas in Secs.~\ref{sec:absorbance_marker_formalism} and \ref{sec:detection_quantum_geometry}.

\begin{figure}[ht]
\begin{center}
\includegraphics[clip=true,width=0.99\columnwidth]{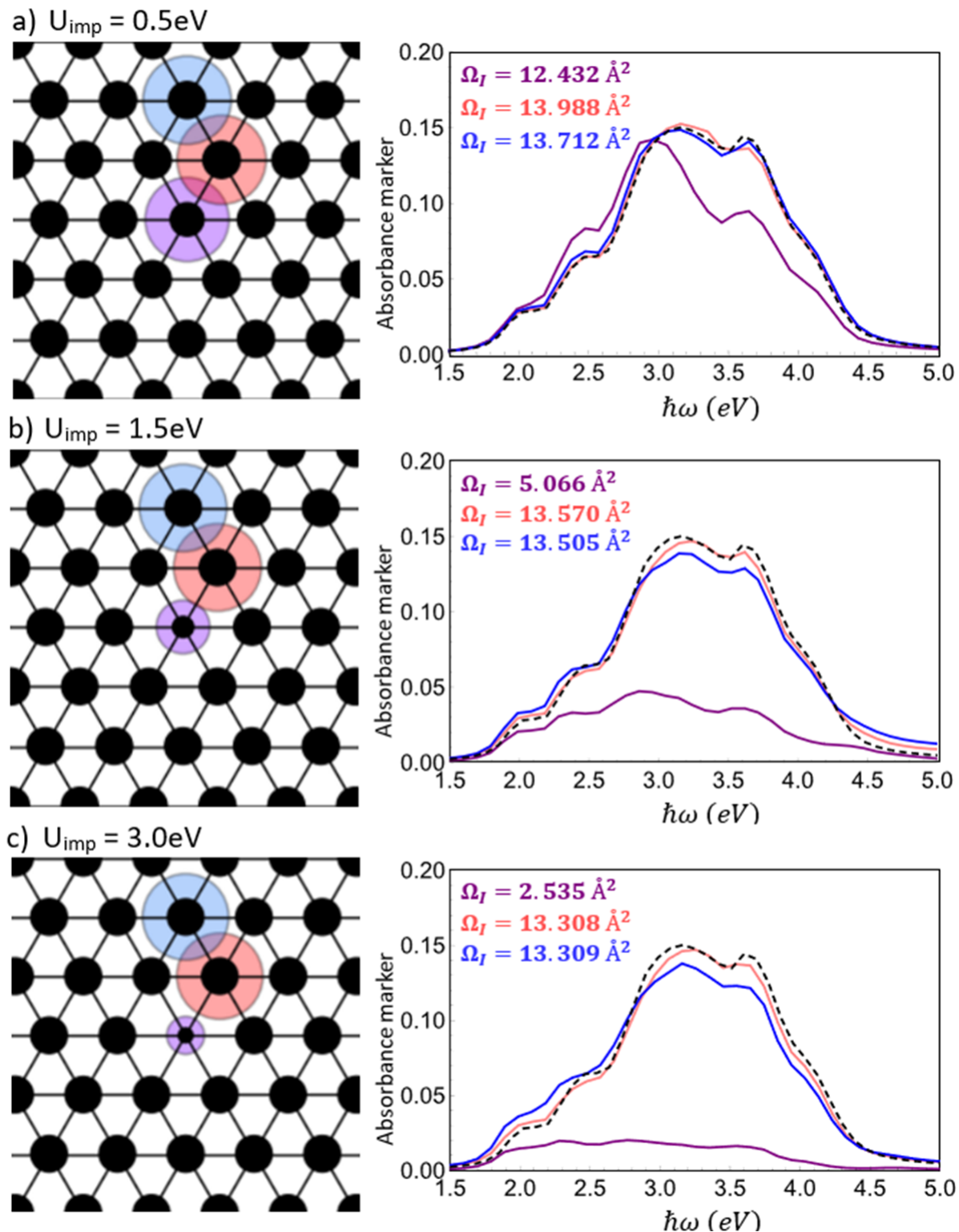}
\caption{The numerical result for a single impurity in WS$_{2}$ with impurity potential (a) $U_{imp}=0.5$eV, (b) $U_{imp}=1.5$eV, and (c) $U_{imp}=3.0$eV. On the left column we plot the particle number on each site (black), and the local spread of Wannier function $\Omega_{I}({\bf r})$ in absolute scale on the impurity site (purple), nearest-neighbor (red) and next-nearest-neighbor (blue) to the impurity site. On the right column we plot the absorbance marker ${\cal O}({\bf r},\omega)$ as a function of frequency on these three sites as a function of frequency using the same color code as the spread of Wannier function. The trace of fidelity marker can be obtained by ${\rm Tr}\,{\cal G}^{d}_{\mu\nu}({\bf r})=\Omega_{I}({\bf r})/A_{\rm cell}$ with $A_{\rm cell}=8.82\AA^{2}$. } 
\label{fig:TMD_single_imp_figure}
\end{center}
\end{figure}

As the first step to clarify the impurity effects, we exam a single impurity in an otherwise homogeneous WS$_{2}$. Figure \ref{fig:TMD_single_imp_figure} shows the numerical results for the simulation on a $14\times 14$ lattice containing a single impurity, with three values of impurity chosen to be smaller than the band gap $U_{imp}=0.5$eV, roughhly equal to the band gap $U_{imp}=1.5$eV, and larger than the band gap $U_{imp}=3$eV. For each value of $U_{imp}$, we plot the real space map of particle number $n({\bf r})$ on each site, the spread of Wannier function $\Omega_{I}({\bf r})$ (equivalent to ${\rm Tr}\,{\cal G}^{d}_{\mu\nu}({\bf r})$ times unit cell area $\approx 8.82\AA^{2}$), and the absorbance marker ${\cal O}({\bf r},\omega)$ as a function of frequency $\omega$ on the few sites near the impurity. From the real space map, one sees a correlation between the reduction of particle number $n({\bf r})$ and the suppression of the spread $\Omega_{I}({\bf r})$, with more suppression at larger impurity potential $U_{imp}$. This is very intuitive, because less particle number means each Wannier orbital is occupied with less electron, and hence a smaller spread. However, the reduction of $n({\bf r})$ and $\Omega_{I}({\bf r})$ recovers quickly as moving away from the impurity site, indicating a very localized impurity effect. Likewisely, the absorbance is significantly reduced at all frequencies at large impurity potential, but quickly recovers as one moves away from the impurity site, similar to the very localized impurity effect on the dielectric and optical properties of 3D semiconductors discovered recently\cite{Chen25_optical_marker}. This localized impurity effect also seems to suggest that when multiple impurities are present and the impurity density is dilute, each impurity acts alone and suppresses the absorbance only in its vicinity, so the reduction of spatially averaged absorbance should be proportional to the impurity density. We shall verify this conjecture in the following section.

\subsection{Multiple impurities and experimental verification in WS$_{2}$}

\begin{figure}[ht]
\begin{center}
\includegraphics[clip=true,width=0.99\columnwidth]{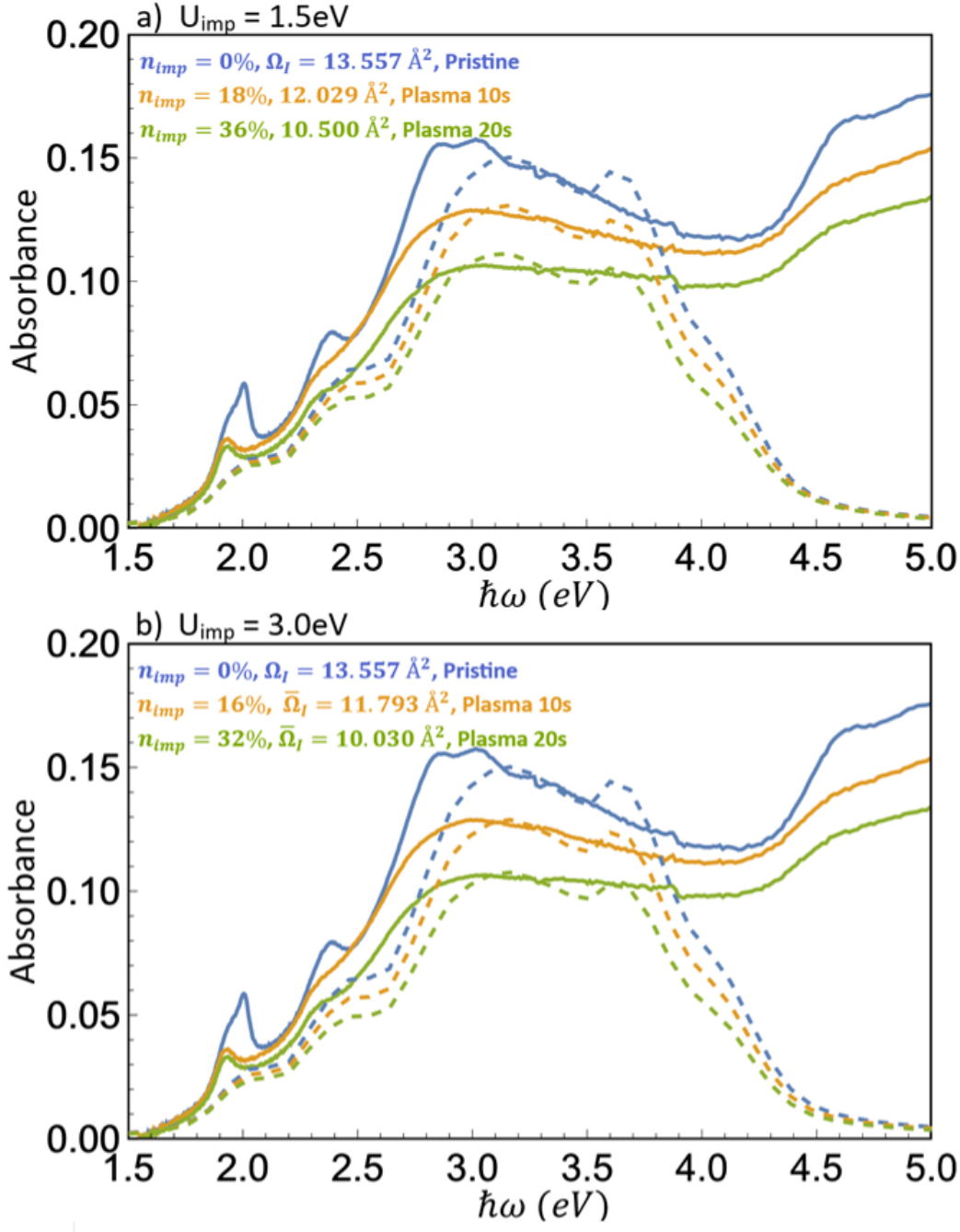}
\caption{The absorbance of plasma-treated WS$_{2}$ (solid lines) and our theoretical fit (dashed lines) using the empirical formula in Eq.~(\ref{absorbance_many_imp_empirical_formula}). In (a), we assume every 10 seconds of plasma exposure produces 18$\%$ of impurity sites, and each impurity has impurity potential $U_{imp}=1.5$eV, while in (b) we assume every 10 seconds of plasma exposure produces 16$\%$ of impurity sites with $U_{imp}=3.0$eV, both yield good agreement with experiments. The average spread of valence band Wannier function $\overline{\Omega_{I}}$ for the $d$-orbitals of W is extracted from each theoretical curve, and the average fidelity marker can be further obtained from $\overline{{\rm Tr}\,{\cal G}_{\mu\nu}^{d}}=\overline{\Omega_{I}}/A_{\rm cell}$. } 
\label{fig:TMD_many_imp_theo_exp}
\end{center}
\end{figure}

In this section, we demonstrate the feasibility of our absorbance marker in capturing the macroscopic scale absorbance measured in disordered 2D semiconductors. For this purpose, we use plasma-treated WS$_{2}$ as an example, where the density of impurities can be controlled by the plasma exposure time. In our experiments, centimeter-scale monolayer WS$_{2}$ was synthesized on fused silica substrates (University WAFER, 500 $\mu$m thickness) via chemical vapor deposition (CVD)\cite{Stand22}. Due to the high optical transparency ($\sim 99\%$) of fused silica, the absorbance of WS$_{2}$ could be readily measured. The optical transparency of pristine monolayer WS$_{2}$ was characterized using a spectrophotometer (Lambda 950 UV-Vis-NIR) with a 2.7 mm diameter aperture, scanning photon energies from 1.5 eV to 5 eV. Subsequently, the sample was subjected to low vacuum air plasma treatment (HARRICK PLASMA plasma cleaner PDC-32G-2) with "LOW" power mode for 5 s to induce partial structural modification. Following plasma exposure, the absorbance of the WS$_{2}$ sample was re-measured. This cycle consisting of absorbance measurement followed by plasma treatment was repeated, and the results of cumulative plasma exposure time of 10 seconds and 20 seconds are presented in Fig.~\ref{fig:TMD_many_imp_theo_exp}.

The absorption spectrum in Fig.~\ref{fig:TMD_many_imp_theo_exp} displays characteristic peaks at approximately 1.9 eV (A exciton) and 2.3 eV (B exciton), which do not contribute to the spread $\Omega_{I}$ since excitons are bosons, so we ignore these peaks in the discussion below\cite{Zhao13,CardenasCastillo24_Wannier_opacity}. For pristine WS$_{2}$, the absorbance caused by interband transition gradually increases from the band gap $\sim 1.5$eV up to the maximum at about $\sim 3$eV, a region where the absorbance is almost entirely contributed from the three $d$-orbitals of W that can be well-described by our theoretical model in Sec.~\ref{sec:tight-binding_model_TMD}. At frequency higher than $\apprge 4$eV, the bands from S atoms start to absorb light, the description of which is beyond our low energy model, so our theoretical investigation aims at fitting the spectrum at frequency lower than $4$eV.

Compared to the pristine WS$_{2}$, Fig.~\ref{fig:TMD_many_imp_theo_exp} shows that the absorbance is progressively suppressed with increasing plasma exposure time, of which we attribute to the removal of S atoms that are usually easier to be removed compared to the W atoms, causing the sample to lose electrons around the removed S atoms. Since our theoretical model only contains W degrees of freedom, we simulate this local loss of electrons by adding a positive impurity potential on certain W sites, which is exactly that described in Sec.~\ref{sec:tight-binding_model_TMD}. In addition, since our theoretical results in Sec.~\ref{sec:tight-binding_model_TMD} indicate a very local suppression of absorbance around the impurity sites, one may model the spatially averaged spread, fidelity marker, and absorbance of the plasma-treated sample in the macroscopic scale defined in Eq.~(\ref{avg_OmegaI_TrG_O}) by the empirical formula 
\begin{eqnarray}
&&\overline{A}=n_{imp}\times A({\bf r}_{imp})+(1-n_{imp})\times A_{hom},
\nonumber \\
&&A=\left\{\Omega_{I},{\rm Tr}\,{\cal G}^{d}_{\mu\nu},{\cal O}(\omega)\right\},
\label{absorbance_many_imp_empirical_formula}
\end{eqnarray}
where $n_{imp}$ is the density of impurity sites assumed to be proportional to the plasma exposure time, $A({\bf r}_{imp})$ is the marker on the impurity site, and $A_{hom}$ is the value in the homogeneous sample. This formula is based on the assumption that at low plasma exposure time where the impurities are dilute, the suppression of absorbance by each impurity is very local as suggested by Fig.~\ref{fig:TMD_single_imp_figure}, so the spatially averaged suppression in the macroscopic scale should be proportional to the density of impurities $n_{imp}$. With this empirical formula, one can fit the experimental data either with a weaker impurity potential and higher impurity density (assuming $U_{imp}=1.5$eV, and an impurity density $n_{imp}=18\%$ is produced by every 10 seconds of plasma exposure) since each impurity would suppress the absorbance moderately, or a stronger impurity potential and lower impurity density (assuming $U_{imp}=3.0$eV, and an impurity density $n_{imp}=16\%$ is produced by every 10 seconds of plasma exposure) since each impurity would suppress the absorbance drastically. To precisely assign the value of $U_{imp}$ under the influence of plasma exposure will require a detailed calculation of the atomic scale kinetics, such as first principle calculations, which is beyond the scope of the present work. Nevertheless, the agreement between the experimental results and our empirical formula in Eq.~(\ref{absorbance_many_imp_empirical_formula}) shown in Fig.~\ref{fig:TMD_many_imp_theo_exp} indicates that within the framework of a simple tight-binding model, our absorbance marker not only describes the local absorbance around the impurity sites, but can also capture the experimentally measured absorbance in the macroscopic scale.


\section{Conclusions}

In summary, we present an absorbance marker that enables the investigation of atomic scale variation of absorbance in 2D semiconductors, as well as the resulting change in quantum geometrical properties and the spread of valence band Wannier functions, caused by disorder. Applying the marker to TMD materials reveals a very localized suppression of absorbance and the spread caused by potential impurities, and as a result the spatially averaged absorbance in the macroscopic scale gradually reduces as the impurity density increases. This theoretical prediction is experimentally verified by the reduced absorbance in WS$_{2}$ caused by plasma exposure, indicating the validity of the absorbance marker. We anticipate that our marker can be ubiquitously applied to a wide variety of realistic 2D semiconductor materials, and new effects on the quantum geometry and the spread of Wannier function caused by various sorts of disorder may be revealed, which await to be further explored. 

\acknowledgements

We acknowledge the financial support from the fellowship for productivity in research from CNPq.

\bibliography{Literatur_abbreviated}

\end{document}